%% file: v8.042.tex
\documentclass{amsart}

\input{settings/settings}
\begin{document}

\title{localization for random CMV matrices}
\author{Xiaowen Zhu}

\begin{abstract}
 We prove Anderson localization (AL) and dynamical localization in expectation (EDL, also known as strong dynamical localization) for random CMV matrices for arbitrary distribution of i.i.d. Verblunsky coefficients.
\end{abstract}
\maketitle

%
\section{Introduction}
\label{sec: intro}

The aim of this paper is to establish Anderson localization and dynamical localization in expectation (see Definition \ref{def: AL} and \ref{def: EDL}) for random CMV matrices with arbitrary distribution.

CMV matrices were introduced by Cantero, L. Moral, L. Vel (\cite{CMV2003}) in 2003 and play an important role in the study of orthogonal polynomials on the unit circle (OPUC). See \cite{Onefoot05}, \cite{5Years07}, \cite{GT2007} for a concise and elegant report of the main results and \cite{OPUC05} for a detailed monograph on this subject.

The study of random CMV matrices was motivated by random Anderson models for Schr\"odinger operators. When the distribution is absolutely continuous, Anderson localization for CMV matrices has been proved in \cite{T1991}, \cite{GN2001}, \cite[Sec. 12.6]{OPUC05} using the spectral averaging method, but these techniques cannot be applied in the singular case. For one-dimensional Anderson model, the first proof that can handle arbitrary randomness was given in \cite{CKM1987}, based on the multi-scale analysis. In 2019, \cite{JZ2019} provided a short proof of Anderson localization and dynamical localization (for the one-dimensional Anderson model with arbitrary distribution) using positive Lyapunov exponents together with uniform large deviation type (LDT) estimates and uniform Craig-Simon results. In 2020, strong dynamical localization was proved in \cite{ZhaoGe20} following this method. In this paper, we exploit the techniques in \cite{JZ2019} and \cite{ZhaoGe20} to prove Anderson localization and strong dynamical localization for random CMV matrices with arbitrary distribution. In particular, our results apply in the singular case. The main novelties of the proof are the large-deviation estimates of determinants with modified boundary conditions (Lemma 4.2) and a streamlined approach to the localization proof in comparison with \cite{JZ2019} and \cite{ZhaoGe20}, so that EDL follows directly from our key statement (Theorem \ref{def: regular}).

It is important to mention that the singular potential random CMV model was also studied in \cite{7Authors19} in 2019 as a close relative of the Anderson model, for which a new proof of localization was also given in \cite{7Authors19}. The CMV proof in \cite{7Authors19} relies on certain results in \cite{Kruger13}. However those contain a significant number of misprints and minor errors (some of those stemming from small misprints in \cite{OPUC05} and \cite{Onefoot05}). Article \cite{7Authors19} inherits those errors, certain steps of the proof in \cite{7Authors19} no longer work as claimed when the relevant expressions are corrected. In particular, a crucial part of the argument of \cite{7Authors19} on the elimination of double resonances does not work as intended although we believe may be corrected. We discuss this in Appendix \ref{App.B}.


As we were completing this paper we learned of \cite{Macera21} where a short proof of Anderson localization for a large class of quasi-one-dimensional operators with singular potentials is presented. It however {doesn't} seem to be applicable to CMV matrices.

Finally, our paper when taken in conjunction with \cite{JZ2019}, \cite{ZhaoGe20}, \cite{Nishant2019}, \cite{NishantEDL2019}, and \cite{Macera21}, illustrates the flexibility of this general scheme for proving localization in random one-dimensional frameworks. Indeed, these techniques provide the most direct route to localization in addition to providing proofs of the strongest known localization results for such models (EDL).



The remainder of the paper is organized as follows:
\begin{itemize}
  \item In Section \ref{sec: modelandresults}, we present the model and the main results (AL and EDL).
  \item In Section \ref{sec: theoremofregular}, we present a key theorem on regularity of Green functions from which AL and EDL are derived.  
  \item In Section \ref{sec: LDTandCS}, we present uniform large deviation theorem (Lemma \ref{lemma: LDT}) and uniform Craig-Simon estimates (Lemma \ref{lemma: uniformCS}).
  \item In Section \ref{sec: pfofthmregular}, we first provide an outline of the proof and prove our key Theorem \ref{thm: regular}.
  \item Finally, Appendix \ref{App.A} provides technical details needed for subsection \ref{ss: GreenFunc} and Appendix \ref{App.B} corrects the errors in the formulas found in \cite{Kruger13}, \cite{7Authors19}, \cite{OPUC05} and \cite{Onefoot05}. It is our hope that these corrections provide clarification for other readers working on CMV matrices.
\end{itemize}

\section{Model and main results}
\label{sec: modelandresults}

\subsection{OPUC}
Let $\eta$ be a probability measure which is supported on an infinite subset of $\partial\DD$ where $\DD$ is the unit disk in $\CC$. Let $\Phi_n(z)$ be the monic polynomial of degree $n$ s.t. 
\begin{equation}
  \label{eq: orthogonal}
  \langle \Phi_m(z), \Phi_n(z)\rangle = \int_{\partial\DD} \overline\Phi_m(z) \Phi_n(z) d\eta(z) = \delta_{mn}, \quad \forall m,n\in\NN.
\end{equation}
The $\Phi_n(z)$'s are called the orthogonal polynomials on the unit circle (OPUC) w.r.t. $\eta$.  Let $\phi_n(z) = \frac{\Phi_n(z)}{\Vert \Phi_n(z)\Vert}$ where $\Vert\cdot\Vert$ is the $L^2(\partial\DD;d\eta)$ norm.

 It is clear that given $\eta$, we can compute $\Phi_n(z)$ and $\phi_n(z)$ inductively from $\Phi_0(z) = \phi_0(z) = 1$. Moreover, there is a recurrence relation for $\Phi_n(z)$ which we state here without proof (see \cite[Theorem 1.5.2]{OPUC05}):

\begin{prop}[Szeg\H{o}'s Recurrence]
  \label{thm: Szego}
  Given $\eta$, there is a sequence of $\alpha_n\in\DD$ s.t. 
    \[
      \begin{split}
      \Phi_{n+1}(z) &= z\Phi_n(z)-\overline{\alpha}_n\Phi_n^*(z)\\
      \Phi_{n+1}^*(z) &= \Phi^*_n(z) - \alpha_n z \Phi_n(z),
      \end{split}
    \]
    where $Q(z)^*:=z^n\overline{Q(1/\bar{z})}$ for polynomials $Q(z)$ of degree $n$. The terms $\{\alpha_n\}_{n=0}^\infty$ are called \textbf{Verblunsky coefficients}. Furthermore, let $\rho_n = (1-|\alpha_n|^2)^{1/2}$. We have
    \[
      \Vert \Phi_n \Vert^2 = \Vert \Phi_n^* \Vert^2 = \rho_n^2\Vert \Phi_{n-1} \Vert^2 = \prod_{k=0}^{n-1}\rho_k^2.
    \]
   Thus, for the normalized $\phi_n$, we have
    \[
      \begin{pmatrix}
        \phi_{n+1}\\ \phi_{n+1}^*
      \end{pmatrix} = \frac{1}{\rho_n}\begin{pmatrix}
        z & -\overline{\alpha}_n\\
        -\alpha_n z & 1
      \end{pmatrix}\begin{pmatrix}
        \phi_n\\\phi_n^*
      \end{pmatrix}.
    \]
\end{prop}

By Szeg\H{o}'s recurrence, each $\eta$ corresponds to a sequence of $\{\alpha_n\}_{n= 0}^\infty\in \DD^{\NN}$. It turns out that this correspondence is bijective (e.g. \cite[Theorem 1.7.11]{OPUC05}).
\begin{prop}[Verblunsky's Theorem]
  There is a bijection between nontrivial (supported on an infinite set) probability measures $\eta$ on $\partial\DD$ and $\{\alpha_n\}_{n=0}^\infty\in\DD^\NN$.
\end{prop}


\subsection{CMV matrices}
\label{ss: CMV}
A CMV matrix is a matrix representation of the multiplication-by-z operator on $L^2(\partial\DD;d\eta)$ w.r.t a basis which is obtained from orthonormalizing the set $\{1,z,z^{-1},z^2,z^{-2},\cdots\}$. It is important to understand the relation between $\eta$ and $\alpha_n$, especially under perturbations. On the one hand, the definition implies that $\eta$ is a spectral measure of the CMV matrix. On the other hand, the CMV matrices can be expressed by the Verblunsky coefficients $\alpha_n$ and $\rho_n = (1-|\alpha_n|^2)^{\frac{1}{2}}>0$ (See \cite[Sec 4.3]{OPUC05} for more details):

\begin{equation}\label{c}
\MC = 
\begin{bmatrix}
  \overline{\alpha_0} & \overline{\alpha_1}\rho_0 & \rho_1\rho_0 \\
  \rho_0 & -\overline{\alpha_1}\alpha_0 & -\rho_1\alpha_0\\
   & \overline{\alpha_2}\rho_1 & -\overline{\alpha_2}\alpha_1 & \overline{\alpha_3}\rho_2 & \rho_3\rho_2\\
   & \rho_2\rho_1 & -\rho_2\alpha_1 & -\overline{\alpha_3}\alpha_2 & -\rho_3\alpha_2\\
   & & & \overline{\alpha_4}\rho_3 & -\overline{\alpha_4}\alpha_3 & \overline{\alpha_5}\rho_4 & \rho_5\rho_4\\
   & & & \rho_4\rho_3 & -\rho_4\alpha_3 & -\overline{\alpha_5}\alpha_4 & -\rho_5\alpha_4\\
   & & & & \ddots & \ddots & \ddots \\
\end{bmatrix}
\end{equation}

We will study a two-sided version of the above matrix. The two-sided version depicted below is called an \emph{extended CMV matrix}. 

\begin{equation}\label{e}
  \ME=
  \begin{bmatrix}
    \ddots & \ddots & \ddots\\
    \overline{\alpha_0}\rho_{-1} & -\overline{\alpha_0}\alpha_{-1} & \overline{\alpha_1}\rho_0 & \rho_1\rho_0 \\
    \rho_0 \rho_{-1} & -\rho_0\alpha_{-1} & -\overline{\alpha_1}\alpha_0 & -\rho_1\alpha_0\\
     & & \overline{\alpha_2}\rho_1 & -\overline{\alpha_2}\alpha_1 & \overline{\alpha_3}\rho_2 & \rho_3\rho_2\\
     & & \rho_2\rho_1 & -\rho_2\alpha_1 & -\overline{\alpha_3}\alpha_2 & -\rho_3\alpha_2\\
     & & & & \overline{\alpha_4}\rho_3 & -\overline{\alpha_4}\alpha_3 & \overline{\alpha_5}\rho_4 & \rho_5\rho_4\\
     & & & & \rho_4\rho_3 & -\rho_4\alpha_3 & -\overline{\alpha_5}\alpha_4 &-\rho_5\alpha_4\\
     & & & & & \ddots & \ddots & \ddots \\
  \end{bmatrix}
\end{equation}
The relationship between $\MC$ and $\ME$ is explained in Remark \ref{rmk: oneside}.

\subsection{Random CMV matrices}
As with the Anderson model, we are interested in the random extended CMV matrix $\ME_\omega$ where $\alpha_n = \omega_n\in \DD$ are i.i.d. random variables with common Borel probability distribution $\mu$ supported on a compact subset $S$ of $\DD$. We assume $\mu$ is non-trivial in the sense that it contains at least two points and as we introduced in the introduction, there are no regularity requirements on $\mu$. Let the probability space be $\Omega=S^{\ZZ}$, with elements $\omega=\{\omega_n\}_{n\in\ZZ}\in\Omega$. Denote $\mu^{\ZZ}$ by $\PP$. Let $\PP_{[m,n]}$ be
$\mu^{[m,n]\cap \ZZ}$ on $\Omega_{[m,n]} := S^{[m,n]\cap \ZZ}$. Hence whenever we write $[m,n]$ in this paper, we mean $[m,n]\cap \ZZ$. Also let $T$ be the shift on $\Omega$, i.e. $(T\omega)_i=\omega_{i-1}$.  Finally, we denote
Lebesgue measure on the unit circle by $m$. 

By the classical ergodicity argument for random operators (e.g. \cite[Chapter 9]{Cycon09}), we see that the spectrum of $\ME_\omega$ is almost surely deterministic, i.e. there is $\Sigma\subset \partial\DD$ s.t. for a.e. $\omega$, $\sigma(\ME_\omega) = \Sigma$. Furthermore, the pure point spectrum, a.c. spectrum and s.c. spectrum are all a.s. deterministic, i.e. $\sigma_{*}(\ME_\omega) = \Sigma_*$, $*\in\{p.p.,a.c.,s.c.\}$.

\subsection{Main results}
We can now introduce our main results.
\begin{definition}[AL]
  \label{def: AL}
  We say $\ME_\omega$ exhibits Anderson localization (AL, also called spectral localization) on $\mathcal{I}$ if for a.e. $\omega$, $\ME_\omega$ has only pure point spectrum in $\mathcal{I}$ and its eigenfunctions $\Psi_\omega(n)$ decay exponentially in $n$.
\end{definition}

\begin{definition}[EDL]
  \label{def: EDL}
  We say $\ME_\omega$ exhibits dynamical localization in expectation (EDL, also known as strong dynamical localization) on $\mathcal{I}$ if there is $C,\eta>0$ s.t. 
  \[
    \sup_{t\in \RR} \EE \left( |\langle \delta_p, e^{-it\ME_\omega}\chi_{\mathcal{I}}(\ME_\omega)\delta_q\rangle |\right) \leq Ce^{-\eta |p-q|}
  \]
  where $\chi_\mathcal{I}$ is the characteristic function of $\mathcal{I}$.
\end{definition}
We will prove in this paper that
\begin{thm}[AL]
  \label{thm: AL}
  There is a set $\mathcal{D}\subset \partial\DD$ which contains at most three points such that, $\ME_\omega$ exhibits AL on any compact interval $\mathcal{I}\subset \partial \DD\setminus \mathcal{D}$.
\end{thm}
\begin{remark}
  \label{rmk: setD}
  The existence of this exceptional set is due to the failure of F\"urstenberg's Theorem (see Subsection \ref{ss: positivityofLy}).
\end{remark}
\begin{thm}[EDL]
  \label{thm: DL}
   There is a set $\mathcal{D}\subset \partial \mathcal{D}$ which contains at most three points s.t. $\ME_\omega$ exibihits EDL on any compact interval $\mathcal{I}\subset \partial\DD\setminus \mathcal{D}$.
\end{thm}

\section{Theorem \ref{thm: regular} implies AL and EDL}
\label{sec: theoremofregular}
Below, we will formulate the key theorem, Theorem \ref{thm: regular}. We then prove AL (Theorem \ref{thm: AL}) and EDL (Theorem \ref{thm: DL}) from it. To do so, we make some preparations in Subsection \ref{ss: DecompofCMV}-\ref{ss: GreenFunc}, state Theorem \ref{thm: regular} in Subsection \ref{ss: StateThms} and prove Theorem \ref{thm: AL} and \ref{thm: DL} in Subsection \ref{ss: prove_AL_and_DL}.

\subsection{Decomposition of CMV matrices}
\label{ss: DecompofCMV}
We start with a decomposition of a CMV matrix which helps us to deal with its more complicated five-diagonal nature.
\label{ss: Decomp}
Let $\alpha_n \in \mathbb{D}$, $\rho_n = (1-|\alpha_n|^2)^{\frac{1}{2}}$. Define the unitary matrix acting on $\ell^2(\{n,n+1\})$ by
\begin{equation}
  \label{eq: decomposition}
  \Theta_n=
  \begin{pmatrix}
    \overline{\alpha_n} & \rho_n\\
    \rho_n & -\alpha_n\\
  \end{pmatrix}.
\end{equation}
Define
\begin{equation}
    \ML=\bigoplus\limits_{n~even} \Theta_n,\quad \MM=\bigoplus\limits_{n~odd} \Theta_n.
\end{equation}
Then one can check directly by computation that the extended CMV matrix satisfies 
\begin{equation}\label{e=lm}
  \ME=\ML\MM.
\end{equation}
By definition of $\Theta_n$, $\alpha_n$ and $\rho_n$, it is easy to see that $\ML$ and $\MM$ are unitary on $\ell^2(\ZZ)$. Thus $\ME$ is also unitary. (More details can be found in \cite[Theorem 4.2.5]{OPUC05}.)

Let $P_\ab : \ell^2(\ZZ)\to \ell^2(\ab)$ be the natural projection. Let $X_\ab  = P_\ab X(P_\ab )^*$
 for $X\in\{\ME,\ML,\MM\}$. Then it is easily verified that
 \begin{equation}
   \label{eq: Eab}
   \ME_\ab  = \ML_\ab \MM_\ab.
 \end{equation} 

\subsection{Modification of the Boundary Conditions}
\label{ss: Bdry}
Notice that $\ME_\ab$, $\ML_\ab$ are not always unitary due to the fact that the ``boundary terms" $\alpha_{a-1}$ and $\alpha_b$ satisfy $|\alpha_{a-1}|<1$ and $|\alpha_b|<1$. Thus we can instead manually create unitary operators by modifying these boundary conditions. Let $\beta,\gamma\in\partial\DD$. Define
\[
  \tilde{\alpha}_n=
  \begin{cases} \alpha_n,\quad &n\neq a-1,b\\
    \beta,\quad &n = a-1\\
    \gamma,\quad &n = b\\
  \end{cases}.
\]
Denote the extended CMV matrix with Verblunsky coefficients $\tilde\alpha_n$ by $\tilde{\ME}$. Then define
\[
  \ME_\ab ^{\beta,\gamma} = P_\ab \tilde{\ME}P_\ab.
\]
$\ML_\ab ^{\beta,\gamma}$ and $\MM_\ab ^{\beta,\gamma}$ are defined correspondingly. Now $\ME_\ab ^{\beta,\gamma}$, $\ML_\ab ^{\beta,\gamma}$ and $\MM_\ab ^{\beta,\gamma}$ are all unitary.
\begin{remark}
Notice that this modification is only a formal modification of the boundary value $|\alpha_{a-1}|<1$ to $|\beta| = 1$ and $|\alpha_b|<1$ to $|\gamma| = 1$. So, all the formulas for $\ME_\ab$ with $\alpha_{a-1}$ and $\alpha_b$ still hold for $\ME^\bg_\ab$ with $\beta$ and $\gamma$. For example, $\ME_\ab ^{\beta,\gamma} = \ML_\ab ^{\beta,\gamma}\MM_\ab ^{\beta,\gamma}$ follows from \eqref{eq: Eab}.
\end{remark}
\begin{remark}\label{rmk: oneside}
  We will use $\ME_\ab^{\beta,\cdot}$, $\ME_\ab^{\beta,\cdot}$ to denote single-sided boundary condition modification. By comparing \eqref{c} and \eqref{e}, it is easy to see that $\mathcal{C}=\ME_{[0,+\infty]}^{-1,\cdot}$. 
\end{remark}

\subsection{Green's functions, Generalized eigenfunctions, Poisson formula}
\label{ss: GreenFunc}
Now we can define the Green's function. Usually it is defined to be $G_\abz = (\ME_\abo - z)^{-1}$. However, since $\ME_\omega$ is five-diagonal, it is more complicated than a Jacobi matrix, and the restriction to $\ab$ is not unitary. Thus we can modify the boundary and rewrite the characteristic function $(\ME_\ab^\bg - z)\Psi = 0$ as $(z(\ML_\ab^\bg)^*-\MM_\ab^\bg)\Psi=0$. Then $A_\abz^\bg := (z(\ML_\ab^\bg)^*-\MM_\ab^\bg)$ is tri-diagonal (see Lemma \ref{matrixA} in the appendix) and it is natural to define the Green's function to be
\[
  G_\abz^\bg = (A_\abz^\bg)^{-1} = \left(z\left(\ML_\ab^\bg\right)^*-\MM_\ab^\bg\right)^{-1}
\]
for $|\beta| = |\gamma| = 1$, $z\notin\sigma(\ME_\ab^\bg)$. 

Exponential decay of the off-diagonal entries of the Green's function turns out to be essential in the study of localization phenomenons. It is closely related to the exponential decay of (generalized) eigenfunctions through Poisson formula. 

\begin{definition}[Generalized eigenvalues and generalized eigenfunctions]
  Fix $\omega$. We call $z_\omega$ a generalized eigenvalue (g.e.) of $\ME_\omega$, if there exists a nonzero, polynomially bounded function $\Psi_\omega(n)$ such that $\ME_\omega\Psi_\omega=z_\omega\Psi_\omega$. We call $\Psi_\omega(n)$ a generalized eigenfunction (g.e.f.).
\end{definition}

\begin{lemma}[Poisson formula]
Let $\Psi$ be a g.e.f. of $\ME$ w.r.t. a g.e. $z$, i.e. $\ME\Psi = z\Psi$. Let $|\beta| = |\gamma| = 1$. Then for $a<x<b$,
  \begin{equation}\label{possion}
    \begin{split}
      \Psi(x)=&-G_{[a,b],z}^{\beta,\gamma}(x,a)
      \begin{cases}
        \Psi(a)(z\bar{\beta}-z\bar\alpha_{a-1})+\Psi(a-1)z\rho_{a-1},\qquad &a~odd\\
        \Psi(a)(\alpha_{a-1}-\beta)-\Psi(a-1)\rho_{a-1},\qquad &a~even\\
      \end{cases}\\
      &-G_{[a,b],z}^{\beta,\gamma}(x,b)
      \begin{cases}
        \Psi(b)(-\bar\alpha_b+\bar\gamma)-\Psi(b+1)\rho_b,\qquad &b~odd,\\
        \Psi(b)(z\alpha_b-z\gamma)+\Psi(b+1)z\rho_b,\qquad &b~even
      \end{cases}
      \end{split}
  \end{equation}
\end{lemma}
We give a proof in Lemma \ref{lemma: poisson} in the appendix.
\medskip

\subsection{Schnol's theorem, Regularity, Key statement}
\label{ss: StateThms}

Recall that Schnol's theorem (see \cite[Theorem 7.1]{Kirsch07}, or \cite[Sec. 2.4]{Cycon09}) says that the spectral measures are supported on the set of g.e.'s. Thus, to show Anderson localization it is enough to show that for a.e. $\omega$, for any g.e. $z_\omega$ of $\ME_\omega$, the corresponding g.e.f. $\Psi_\omega$ decays exponentially, because this would imply that each g.e. is indeed an eigenvalue, so $\ME_\omega$ has only pure point spectrum. 

Thus for a g.e.f. $\Psi_\omega$ which is polynomially bounded, if we can show the Green's function $|G_{[n+1,3n+1],\omega,z}^\bg(2n+1,n+1)|$ and $|G_{[n+1,3n+1],\omega,z}^\bg(2n+1,3n+1)|$ are exponentially small, then $|\Psi_\omega(2n+1)|$ will decay exponentially due to the Poisson formula. This idea inspires us to define regularity as follows:
\begin{definition}[Regularity]
  \label{def: regular}
  Let $\beta$, $\gamma\in \partial\DD$. For fixed $\omega$, $z\notin \sigma(\ME_\abo^\bg)$, $c>0, n\in\ZZ$, we say $x\in\ZZ~$ is $(c,n,\omega,z)$-regular, if
 \[
   \begin{split}
    &|G_{[x-n,x+n],\omega,z}^{\beta,\gamma}(x,x-n)| \leq e^{-cn}\\
    &|G_{[x-n,x+n],\omega,z}^{\beta,\gamma}(x,x+n)| \leq e^{-cn}     
   \end{split}
 \]
 Otherwise, we call it $(c,n,\omega,z)$-singular.
\end{definition}

\subsection{Proof of AL and EDL}
\label{ss: prove_AL_and_DL}
  We can now formulate our key statement:
\begin{thm}
  \label{thm: regular}
  There is a set $\mathcal{D}\subset \partial\DD$ which contains at most three points such that, for any compact interval $\mathcal{I}\subset \partial\DD\setminus \mathcal{D}$, if we let $\nu := \inf\limits_{z\in \mathcal{I}} \gamma(z)>0$, then for any $0<\epsilon<\nu/2$, there is $N = N(\epsilon),\eta = \eta(\epsilon)>0$ s.t. $\forall n>N$, $\forall x\in\ZZ$, there is a subset $\oxn\subset \Omega_{[x-n,x+n]}$ s.t. 
    \begin{enumerate}
    \item $\PP(\oxn)\geq 1 - e^{-\eta (2n+1)}$. 
    \item $\forall \omega\in \oxn$, either $x$ or $x+2n+1$ is $(\gamma(z)-2\epsilon,n,\omega, z)$-regular for any $z\in \mathcal{I}$.
  \end{enumerate}
\end{thm}
  This Theorem will be proved in the next two sections. We first show Theorem \ref{thm: regular} implies Theorem \ref{thm: AL} and Theorem \ref{thm: DL} before proving Theorem \ref{thm: regular}.
\begin{proof}[Proof of Theorem \ref{thm: AL} (AL)]
    Find $\mathcal{D}$, $\mathcal{I}$, $\nu$ from Theorem \ref{thm: regular}. For any $0<\epsilon<\nu/2$, find $N(\epsilon)$, $\eta(\epsilon)$, $\oxn$ from Theorem \ref{thm: regular}. For any $x\in\ZZ$, since $\sum\limits_n \PP\left((\oxn)^c\right)<\infty$, by the Borel-Cantelli Lemma, for a.e. $\omega$, eventually either $x$ or $x+2n+1$ is $(\gamma(z)- 2\epsilon,n,\omega,z)$-regular. 
    
    On the other hand, for a.e. $\omega$, take any g.e. $z\in \mathcal{I}$. Let $\Psi_\omega(m)$ be  the corresponding g.e.f.. WLOG assume $\Psi_\omega(x) \neq 0$. Thus by Lemma \ref{lemma: poisson}, we claim that for such $x$, $\omega$, $z$ and $\Psi_\omega$, $x$ is eventually $(\gamma(z) - 2\epsilon, n,\omega,z)$-singular. For if $x$ is $(\gamma(z) - 2\epsilon, n,\omega,z)$-regular infinitely often, then $\Psi_\omega(x) = 0$.

    Since $x$ is eventually $(\gamma(z)- 2\epsilon,n,\omega,z)$-singular,  $x+2n+1$ is $(\gamma(z)- 2\epsilon,n,\omega,z)$-regular. Thus $\Psi_\omega(x+2n+1)$ decays exponentially as $n\to \infty$. A similar argument applies to  $\Psi_\omega(x+2n+2)$. Therefore, for a.e. $\omega$, all of the g.e.f.'s $\Psi_\omega(n)$ decay exponentially.
\end{proof}

\begin{proof}[Proof of Theorem \ref{thm: DL} (EDL)]
By Theorem \ref{thm: AL}, for a.e. $\omega$, there is an orthonormal basis $\{\Psi_{k,\omega}\}$ of eigenfunctions of $\ME_\omega$. Denote the corresponding eigenvalues by $z_{k,\omega}$. Define the localization center as the left-most $c_{k,\omega}\in\ZZ$ s.t. 
\[
  |\Psi_{k,\omega}(c_{k,\omega})| = \max\limits_{n\in\ZZ} |\Psi_{k,\omega}(n)|.
\]
We will employ the following Lemma from \cite{JK2012} which provides a sufficient condition for EDL:
\begin{lemma}[\cite{JK2012}]
  \label{lemma: EDLcriteria}
  If there are $\tilde{C}>0$, $\tilde{\gamma}>0$, s.t. for any $x,y\in\ZZ$ 
  \begin{equation}
    \label{eq: criteriaofEDL}
    \EE(\sum\limits_{k:c_{k,\omega} = y} |\Psi_{k,\omega}(x)|^2 ) \leq \tilde{C} e^{-\tilde{\gamma}|x-y|},
  \end{equation}
      Then there are $C>0$, $\gamma>0$, s.t. 
  \[
    \sup\limits_{t\in\RR} \EE(|\langle \delta_p,e^{it\ME_\omega}\chi_I(\ME_{\omega}))\leq C(|x-y|+1) e^{-\gamma|x-y|}.
  \]
\end{lemma}
 Thus, we only need to show \eqref{eq: criteriaofEDL}. To do so, observe that since $c_{k,\omega}$ is where $\psi_{k,\omega}(x)$ takes the supremum value. It is naturally $(\gamma(z) - 2\epsilon, n,\omega,z)$-singular for any $n$ s.t. $e^{-(\gamma - 2\epsilon)n}<\frac{1}{2}$. By Theorem \ref{thm: regular}, if $|x - y| = |x - c_{k,\omega}|>2N$, then $x$ is $(\gamma(z) - 2\epsilon, n,\omega,z)$-regular. Thus we have 
 \[
  |\Psi_{k,\omega}(x)|\leq 2|\Psi_{k,\omega}(y)|e^{-(\gamma(z) - 2\epsilon)(|x - y|)}, \quad \forall |x - y|>2N, \omega\in\Omega_{\Lambda_{2N}(y)}.
 \]
Since $\Psi_{k,\omega}$ is an orthonormal basis, by Bessel's Inequality, we have
\[
  \sum\limits_{k:c_{k,\omega} = y} |\Psi_{k,\omega}(x)|^2 \leq 4\sum\limits_{k:c_{k,\omega} = y} |\Psi_{k,\omega}(y)|^2e^{-(\nu - 2\epsilon)|x - y|} \leq 4e^{-(\nu - 2\epsilon)|x - y|}.
\]
 Thus if $|x - y|>2N$, we have
 \begin{equation}
  \label{eq: EDLcase1}
  \begin{split}
    \EE(\sum\limits_{k:c_{k,\omega} = y} |\Psi_{k,\omega}(x)|^2 )  &\leq \int_{\Omega_{\Lambda_{2N}(y)}} \sum\limits_{k:c_{k,\omega} = y}|\Psi_{k,\omega}(x)|^2 d\PP(\omega) + \int_{(\Omega_{\Lambda_{2N}(y)})^c} \sum\limits_{k:c_{k,\omega} = y}|\Psi_{k,\omega}(x)|^2 d\PP(\omega)\\
    &\leq  1*4e^{-(\nu -2\epsilon)|x - y|} + e^{-\eta|x - y|}*1\\
    &\leq 5e^{-\tilde{\gamma}|x - y|}
   \end{split}
 \end{equation}
  with $\tilde{\gamma} = \min \{\nu - 2\epsilon, \eta\}$.
 If $|x - y|\leq 2N$, since $\Psi_{k,\omega}$ is orthonormal basis, we have 
 \begin{equation}
   \label{eq: EDLcase2}
   \sum\limits_{k:c_{k,\omega} = y} |\Psi_{k,\omega}(x)|^2\leq 1, \quad \text{thus} \quad \EE(\sum\limits_{k:c_{k,\omega} = y} |\Psi_{k,\omega}(x)|^2 )\leq 1.
 \end{equation}
  Since there are finitely many $x$'s in the case when $|x - y|\leq 2N$, combining \eqref{eq: EDLcase1} and \eqref{eq: EDLcase2}, we see that there is $\tilde{C}$, s.t. for any $x$
 \[
  \sum\limits_{k:c_{k,\omega} = y} |\Psi_{k,\omega}(x)|^2\leq \tilde{C} e^{-\tilde{\gamma}|x - y|}
 \]
 Having shown \eqref{eq: criteriaofEDL}, EDL follows by Lemma \ref{lemma: EDLcriteria}.
\end{proof}

\section{Uniform LDT Estimates and Uniform Craig-Simon Results}
\label{sec: LDTandCS}
 In this section, we introduce the uniform large-deviation-type estimates (uniform LDT) and uniform Craig-Simon results which are preliminary results needed for the proof of Theorem \ref{thm: regular}. We begin by connecting the Green's function with determinants of box-restrictions, transfer matrices and Lyapunov exponents. 
 
 \subsection{Determinants with boundary conditions}
Let
  \begin{equation}
    \label{eq: def_of_determinants}
    \begin{split}
      &\mathcal{P}_\aboz^\bg := \det(z-\ME_\abo^\bg) = \det(A_\ab^\bg),\\
      &P^\bg_\aboz := (\rho_{a-1}\cdots\rho_{b})^{-1}\mathcal{P}^\bg_\aboz.
      \end{split}
  \end{equation}
  If $a> b$, let $P_{[a,b],\omega,z}^\bg = 1$. Note that although we have modified the boundary conditions in $\mathcal{P}_\aboz^\bg$, we keep $\rho_{a - 1}$ and $\rho_b$ unchanged in the second formula above. Moreiver, $P_{[a,b],\omega,z}^{\beta,\cdot}$ and $P_{[a,b],\omega,z}^{\cdot,\gamma}$ are defined similarly.

By Cramer's rule, we have 
  \begin{equation}
    \label{eq: GreenFunc}
    \begin{split}
      \left\vert G_{[a,b],\omega,z}^{\beta,\gamma}(x,y)\right\vert &= \frac{|\mathcal{P}_{[a,x-1],\omega,z}^{\beta, \cdot} \mathcal{P}_{[y+1,b],\omega,z}^{\cdot, \gamma}|}{\mathcal{P}_\aboz^\bg}\prod\limits_{k = x}^{y-1}\rho_k\\
      &= \frac{\left\vert P_{[a,x-1],\omega,z}^{\beta,\cdot}P_{[y+1,b],\omega,z}^{\cdot,\gamma}\right\vert}{\left\vert P_{[a,b],\omega,z}^{\beta,\gamma}\right\vert},\quad a\leq x\leq y\leq b\\
    \end{split}
  \end{equation}

  \subsection{Transfer Matrix and Lyapunov Exponents}
Recall by Theorem \ref{thm: Szego},  
\[
  \begin{pmatrix}
    \phi_{n+1}(z)\\ \phi_{n+1}^*(z)
  \end{pmatrix} = \frac{1}{\rho_n}\begin{pmatrix}
    z & -\overline{\alpha_n}\\
    -\alpha_n z & 1
  \end{pmatrix}\begin{pmatrix}
    \phi_n(z)\\\phi_n(z)^*
  \end{pmatrix}
\]
Denote 
\[
  S_z(\alpha)=\frac{1}{\rho_\alpha}
  \begin{pmatrix}
    z &-\overline{\alpha}\\
    -\alpha z & 1\\
  \end{pmatrix},
\]
Let $T_\ab = S_z(\alpha_b)\cdots S_z(\alpha_a)$ be the transfer matrix, then 
\[
  \begin{pmatrix}
    \phi_{b+1}(z)\\ \phi_{b+1}^*(z)
  \end{pmatrix} = T_\ab\begin{pmatrix}
    \phi_a(z)\\ \phi_a^*(z)
  \end{pmatrix}.
\]
By \cite[Theorem 1]{Wang18} together with Remark \ref{rmk: oneside}, we have:
\begin{equation}
  T_\ab =\frac{1}{\rho_a\cdots\rho_b}
  \begin{bmatrix}
    z\mathcal{P}_{[a+1,b],z} & \mathcal{P}_{[a,b],z}^{-1,\cdot}-z\mathcal{P}_{[a+1,b],z}\\
    z(\mathcal{P}_{[a,b],z}^{-1,\cdot}-z\mathcal{P}_{[a+1,b],z})^* & (\mathcal{P}_{[a+1,b],z})^*\\
  \end{bmatrix}\\
\end{equation}
or
\begin{equation}\label{*}
    T_\ab =
  \begin{bmatrix}
    zP_{[a+1,b],z} & \rho_{a-1}P_{[a,b],z}^{-1,\cdot}-zP_{[a+1,b],z}\\
    z(\rho_{a-1}P_{[a,b],z}^{-1,\cdot}-zP_{[a+1,b],z})^* & (P_{[a+1,b],z})^*\\
  \end{bmatrix}
\end{equation}
where $Q(z)^*=z^n\overline{Q(1/\bar{z})}$ if $Q(z)$ is a polynomial of degree n.

Note that $\frac{1}{\sqrt{z}}S_z(\alpha)\in SU(1,1)$, where 
\[
    SU(1,1) = \left\{\begin{pmatrix}
      u & v \\ \bar{v} & \bar{u}
    \end{pmatrix}: u,v\in \CC, |u|^2-|v|^2 = 1\right\},
\]
and for $\sqrt{z}$, we take the principal branch. 
Note also that $SU(1,1) = Q^{-1}\cdot SL(2,\RR)\cdot Q$, where $ Q = -\frac{1}{2}\begin{pmatrix}
    i & 1 \\ 1 & i
\end{pmatrix}$. Thus, the definitions of Lyapunov exponents for $SL(2,\RR)$-cocycles and the corresponding properties (positivity and continuity, large deviation and subharmonicity results) generalize to $SU(1,1)$-cocycles. Moreover, $\Vert S_z(\alpha)\Vert = \Vert \frac{1}{\sqrt{z}} S_z(\alpha)\Vert$ when $|z| = 1$. Thus, when the $\alpha_n$'s are i.i.d., by Kingman's subadditive theorem (\cite{Kingman1973}), the Lyapunov exponent $\gamma(z)$ is well-defined:
\begin{equation}\label{gamma}
  \gamma(z)=\lim_{n\to\infty}\frac{1}{n}\int_0^1 \log\Vert
  T_{[0,n],\omega,z}\Vert
  d\mathbb{P}(\omega)=\lim_{n\rightarrow\infty}\frac{1}{n} \log\Vert
  T_{[0,n],\omega,z}\Vert, \quad \text{~a.e.~} \omega.
\end{equation}

\subsection{Positivity and Continuity of Lyapunov Exponent}
\label{ss: positivityofLy}
By F\"urstenberg Theorem, random Sch\"odinger operators have positive Lyapunov exponent: $\gamma(z)>0$ for any $z\in\RR$. At the same time, random CMV matrices may have an exceptional set $\mathcal{D}\subset \partial\DD$ which contains at most three points s.t. $\gamma(E)>0$ on $\partial\DD\setminus\mathcal{D}$. In fact, depending on the support of $\mu$, either $\mathcal{D} = \emptyset$ or $\mathcal{D} = \{1,-1\}$ or $\mathcal{D} = \{1,\theta_0, \overline{\theta_0}\}$, for some $\theta_0\in\partial\DD$. The reason is, roughly speaking, the positivity get destroyed when $S_z(\alpha_i)$ and $S_z(\alpha_j)$ have a common invariant measure. This would happen only if $\alpha_i$, $\alpha_j$ and $z$ satisfy certain algebraic conditions which characterize the exceptional set. See \cite[Theorem 12.6.3. and 10.4.18]{OPUC05} for more details. 

Continuity of Lyapunov exponents on $\mathcal{I}\subset \partial\DD\setminus \mathcal{D}$ can be proved using the general method (e.g. \cite[Sec.V.4.2]{CarmonaLacroix12}, \cite{BougerolLacroix2012}) originally developed by F\"urstenberg and Kifer \cite[Theorem B]{FurstenKifer83} for self-adjoint random matrices, which by conjugation, extend to $SU(1,1)$ random matrices naturally. We also refer to \cite[Sec.2]{7Authors19}, \cite[Sec. 7]{HamzaStolz07} for a review of the proof of continuity of Lyapunov exponents.

We fix an interval $\mathcal{I}\subset \partial\DD\setminus \mathcal{D}$ from now on and denote $\nu := \inf\limits_{z\in\mathcal{I}} \gamma(z)>0$.

\subsection{Uniform Large-deviation-type estimates}
We can now introduce the uniform large-deviation-type estimates, a crutial component of the proof of Theorem \ref{thm: regular}. These LDT type estimates for $\Vert T_{[a,b],\omega,z}\Vert$ were proved in \cite{Lepage82}. Here we use a matrix-entry version from \cite[Theorem 5]{Tsay99}. The result was proved for $SL(2,\RR)$-cocycle. Here by conjugation, we rewrite it for our $SU(1,1)$-cocycle $T_{[a,b],\omega,z}$. So under the same assumption for positivity and continuity of Lyapunov exponent, which, in particular, holds for any compact interval $\mathcal{I}\subset \partial\DD$, we have the following lemma:

\begin{lemma}[ ``uniform-LDT'']
  \label{ldt lemma}
  Given a compact interval $\mathcal{I}\subset \partial\DD\setminus \mathcal{D}$. For any $\epsilon>0$, there exists $\eta=\eta(\epsilon, \mathcal{I})$, $N=N(\epsilon,\mathcal{I})>0$, such that
    \begin{equation}\label{eqLDT2}
      \mathbb{P}\left\{ \omega:\left\vert 
      \frac{1}{b-a+1} \log\vert \langle T_{[a,b],\omega,z}u,v\rangle\vert-\gamma(z) \right\vert\geq\epsilon
       \right\} \leq e^{-\eta (b-a+1)}
     \end{equation}
     for any $b-a>N$, for any unit vector $u$, $v$ and any $z\in \mathcal{I}$.
\end{lemma}
Thus for our model, we have 
\begin{lemma} 
  \label{lemma: LDT}
  Given a compact interval $\mathcal{I}\subset \partial\DD\setminus \mathcal{D}$. For any $\epsilon>0$, there is an $\tilde{\eta}=\tilde{\eta}(\epsilon,\mathcal{I}),\tilde{N}_1=\tilde{N}_1(\epsilon,\mathcal{I})>0$ s.t. 
  \begin{align}
      \mathbb{P}&\left\{ \omega:\left\vert \frac{1}{b-a+1} \log\vert P_{[a,b],\omega,z}^{-1,\cdot}\vert-\gamma(z) \right\vert\geq\epsilon \right\} \leq e^{-\eta (b-a+1)}\label{eq: LDT1}\\
      \mathbb{P}&\left\{ \omega:\left\vert \frac{1}{b-a+1} \log\vert P_{[a,b],\omega,z}^{\cdot,1}\vert-\gamma(z) \right\vert\geq\epsilon \right\} \leq e^{-\eta (b-a+1)}\label{eq: LDT2}\\
      \mathbb{P}&\left\{ \omega:\left\vert \frac{1}{b-a+1} \log\vert P_{[a,b],\omega,z}^{-1,1}\vert-\gamma(z) \right\vert\geq\epsilon
     \right\} \leq e^{-\eta (b-a+1)}\label{eq: LDT3}
    \end{align}  
    for every $b-a>\tilde{N}_0$, and any $z\in\mathcal{I}$.
\end{lemma}

\begin{proof}
  First recall that $\alpha_k$ is supported on a compact subset of $\DD$, $\rho_k = \sqrt{1 - |\alpha_k|^2}$. Thus there is $\delta>0$ s.t. 
  \begin{equation}
    \label{eq: bounded}
    \begin{cases}
      |\alpha_k|\leq 1 - \delta<1,\\
      0<\delta\leq |\rho_k|\leq  1 - \delta<1.
    \end{cases}
  \end{equation}
  \eqref{eq: LDT1}, \eqref{eq: LDT2} and \eqref{eq: LDT3} above each require seperate consideraions. 
  
  To prove \eqref{eq: LDT1}, let $u= (1,0)^T$, $v = (1,1)^T$ in Lemma \ref{ldt lemma}. By \eqref{*}, the $zP_{[a+1,b],z}$ term cancels and we get 
    \[
      \mathbb{P} \left\{ \omega:\left\vert \frac{1}{b-a+1} \log\vert \rho_{a-1}P_{[a,b],\omega,z}^{-1,\cdot}\vert-\gamma(z) \right\vert\geq\epsilon \right\} \leq e^{-\eta (b-a+1)}.
    \]
    By \eqref{eq: bounded}, $\rho_{a-1}$ can be absorbed by large enough $b-a+1$ and a modified $\epsilon$.

    As for \eqref{eq: LDT2}, the inequality follows from \eqref{eq: LDT1} by setting $\widetilde{\alpha}_j=-\overline{\alpha}_{a+b-1-j}$ for $a-1\leq j\leq b$ and observing that $P_{[a,b],z}^{\cdot,1}(\widetilde{\alpha}_j) = P_{[a,b],z}^{-1,\cdot}(\alpha_j)$. 

     Lastly, to prove \eqref{eq: LDT3}, notice that by \eqref{eq: def_of_determinants} and Lemma \ref{matrixA}, we have 
    \[
        \left|P_{[a,b],z}^{-1,1}\right|= \frac{\left|\det(A_{[a,b],z}^{-1,1})\right|}{\rho_{a-1} \cdots \rho_b} =\frac{\left|A_{b,b}^{\alpha_b=1}\mathcal{P}_{[a,b-1]}^{-1,\cdot}-A_{b,b-1}\prod_{n=a}^{b-1}A_{n,n+1}\right|}{\rho_{a-1} \cdots \rho_b}.
    \]
     And by Lemma \ref{matrixA} and \eqref{eq: bounded}, we have $\delta<|A_{b,b}^{\alpha_b = 1}|<2$. Thus by Lemma \ref{matrixA}, the first and second terms are bounded respectively by
    \[
      \begin{split}
        &\frac{\delta\left|P_{[a,b-1]}^{-1,\cdot}\right|}{\delta}  \leq \frac{\left|A_{b,b}^{\alpha_b = 1}\mathcal{P}_{[a,b-1]}^{-1,\cdot}\right|}{\rho_{a-1}\cdots\rho_b} = \frac{\left|A_{b,b}^{\alpha_b = 1}P_{[a,b-1]}^{-1,\cdot}\right|}{\rho_b}\leq \frac{2\left|P_{[a,b-1]}^{-1,\cdot}\right|}{\delta} \\
        &\delta\leq \frac{\left|A_{b,b-1}\prod_{n=a}^{b-1}A_{n,n+1}\right|}{\rho_{a-1} \cdots \rho_b} = \frac{\rho_{b-1}\prod_{n = a}^{b-1}\rho_n}{\rho_{a-1}\cdots \rho_b} = \frac{\rho_{b-1}}{\rho_{a-1}\rho_b}\leq \frac{1}{\delta^2}.
      \end{split}
    \]
    Hence,  
    \[
      \left|P_{[a,b-1]}^{-1,\cdot}\right| - \frac{1}{\delta^2} \leq \left|P_{[a,b-1]}^{-1,1}\right| \leq \frac{2}{\delta} \left|P_{[a,b-1]}^{-1,\cdot}\right| + \delta.
    \]
    and the third inequality follows.
 \end{proof}

\subsection{``Bad sets" and Singularity}
To simplify the notation, we introduce  ``bad sets" and use them to characterize ``singularity'' in Defintion \ref{def: regular}. Denote
\begin{equation}\label{B+}
  \begin{split}
    B_{[a,b],\epsilon}^{{\beta,\gamma},+} =\left\{(\omega,z)\in \mathcal{I}\times \Omega: |P_{[a,b],\omega,z}^{\beta,\gamma}|\geq e^{(\gamma(z)+\epsilon)(b-a+1)}\right\}\\
    B_{[a,b],\epsilon,}^{{\beta,\gamma},-} =\left\{(\omega,z)\in \mathcal{I}\times \Omega: |P_{[a,b],\omega,z}^{\beta,\gamma}|\leq e^{(\gamma(z)-\epsilon)(b-a+1)}\right\}\
  \end{split}
\end{equation}
Let $B_{[a,b],\epsilon,z}^{{\beta,\gamma},\pm}$ and $B_{[a,b],\epsilon,\omega}^{{\beta,\gamma},\pm}$ be the $z$ and $\omega$ sections of $B_{[a,b],\epsilon}^{{\beta,\gamma},\pm}$. Let $B_{[a,b],\epsilon,*}^{\beta,\gamma}=B_{[a,b],\epsilon,*}^{\beta,\gamma,+}\cup B_{[a,b],\epsilon,*}^{\beta,\gamma,-}$. All of these sets have corresponding definitions for the single-sided boundary case. Thus, \eqref{eq: LDT1}, \eqref{eq: LDT2}, \eqref{eq: LDT3} can be rewritten as 
\begin{equation}\label{ldt}
    \mathbb{P}(B_{[a,b],\epsilon,z}^*)\leq e^{-\eta(b-a+1)}
\end{equation}
where $*$ can be any of the three kinds of boundary conditions $\bg$ or $\beta,\cdot$, or $\cdot,\gamma$. 

We can characterize singular points using the bad sets:
\begin{lemma}
  \label{lemma: singular}
For any $\epsilon<\nu/2$, for $n \geq 2$, if $x$ is $(\gamma(z)-2\epsilon,n,\omega,z)$-singular, then
\[(\omega,z)\in B_{[x-n,x+n],\epsilon}^{\beta,\gamma,-}\cup B_{[x-n,x-1],\epsilon}^{\beta,\cdot,+}\cup B_{[x+1,x+n],\epsilon}^{\cdot,\gamma,+}
\]
\end{lemma}

\begin{proof}
The result follows imediately from the definition of singularity and \eqref{eq: GreenFunc}.
\end{proof}
\subsection{Uniform Craig-Simon results}
We will also use a uniform version of Craig-Simon's results. The Craig-Simon estimates \cite{CraigSimon83} are a general subharmonicity upper bound estimate. It is extended in \cite[Theorem 5.1]{JZ2019} to the uniform version. See \cite[Section 5]{JZ2019} for more details. 

\begin{lemma}[Uniform Craig-Simon]
  \label{lemma: uniformCS}
  Let $\ME_\omega$ satisfy uniform-LDT condition in Lemma \ref{ldt lemma}. Then for any $\epsilon$, there is $\tilde{\eta}=\tilde{\eta}(\epsilon),\tilde{N}_2=\tilde{N}_2(\epsilon)>0$ s.t. for any $x\in\ZZ$, $n>N_1$, there is $\tilde{\Omega}_{x,n}$ s.t. 
  \begin{enumerate}
    \item $\PP(\tilde{\Omega}_{x,n})\geq  1 - ne^{-\eta(n+1)}$,
    \item for any $\omega\in\tilde{\Omega}_{x,n}$, we have for every $z\in\mathcal{I}$.
        \[
            \max\{|P_{[x+1,x+n],z}^{\beta,\cdot}|,|P_{[x-n,x-1],z}^{\cdot,\gamma}|\}\leq e^{(\gamma(z)+\epsilon)(n+1)}, \quad \text{i.e.}
        \]
 \begin{equation}
    \label{eq: uniformCS}
    (\omega,z) \notin B_{[x+1,x+n],\epsilon}^{\beta,\cdot,+}\cup B_{[x-n,x-1],\epsilon}^{\cdot,\gamma,+}.
  \end{equation}
  \end{enumerate}
\end{lemma}
\begin{proof}
  The deterministic result is a direct reformulation of \cite[Theorem 5.1]{JZ2019}, while the probabilistic results can be extracted from the last line on Page 9 in \cite{JZ2019}.
\end{proof}
\begin{remark}
  We mention in particular that $\tilde{\eta}$ in Lemma \ref{lemma: LDT} and Lemma \ref{lemma: uniformCS} for the same $\epsilon$ are the same. In fact, the $\tilde{\eta}$ in Lemma \ref{lemma: uniformCS} comes from applying \ref{lemma: LDT}.  (See \cite{JZ2019}).
\end{remark}

\section{Proof of Theorem \ref{thm: regular}}
\label{sec: pfofthmregular}
We will prove Theorem \ref{thm: regular} in this section. Heuristically, Theorem \ref{thm: regular} says, with high probability, one of two points will be regular if they are far enough from each other. The idea is that with high probability, if $x$ is a $(\gamma(z) - 2\epsilon,n,\omega,z)$-singular point, then $z$ will be exponentially close to $\sigma(\ME_{[x-n,x+n],\omega})$. We will denote this set by $\oa$. So, if we have two far away singular points $x$, $y$, then $\sigma(\ME_{[x-n,x+n],\omega})$ and $\sigma(\ME_{[x-n,x+n],\omega})$ are also exponentially close to the same $z$. However, we can also show that with high probability $\sigma(\ME_{[x-n,x+n],\omega})$ and $\sigma(\ME_{[x-n,x+n],\omega})$ cannot be exponentially close. We will denote this set by $\ob$. Then $\oa\cap \ob$ will be the set of high probability where one of these two points must be regular.

For convenience, we will omit $\omega,z$ from the subscript of $T_{[a,b],\omega,z}$, $P_{[a,b],\omega,z}^*$, $G_{[a,b],\omega,z}^*$ and $A_{[a,b],\omega,z}$ in this section unless it is necessary.

\medskip

\subsection{The first set $\oa$}
\label{ss: Omega_1}
As we mentioned above, we choose $\oa$ s.t. singularity implies exponential closeness to the  spectrum:
\begin{lemma}
  \label{lemma: oa}
  For any $0<\epsilon<\nu$, there are $\eta_1 = \eta_1(\epsilon)$ , $N_1 = N_1(\epsilon_1)$  s.t. for any $n>N_1$, $x\in\ZZ$, $0<\delta<\eta_1$, there is $\oxn^{(1)} = \oxn^{(1)}(\delta)$, s.t. 
  \begin{enumerate}
    \item $\PP(\oxn^{(1)})\geq 1 - m(\mathcal{I})e^{-(\eta_1 - \delta)(2n+1)} - ne^{-\eta_1(2n+1)}$, 
    \item For $\omega\in\oxn^{(1)}$, if $x$ is $(\gamma(z) - 2\epsilon, n,\omega,z)$-singular, then 
    \[
      \operatorname{dist}(z,\sigma(\ME_{[x-n,x+n],\omega}))\leq e^{-\delta(2n+1)}.
    \]
  \end{enumerate}
\end{lemma}

\begin{proof}
  Fix any $0<\epsilon<\nu/2$. Let $\tilde{\eta}(\epsilon)$, $\tilde{N}_1(\epsilon)$ be as in Lemma \ref{lemma: LDT}. Let $\tilde{N}_2(\epsilon)$, $\tilde{\Omega}_{x,n}$ be as in Lemma \ref{lemma: uniformCS}. Then let $\eta := \tilde{\eta}$, $N := \max\{N_1,N_2\}$, and
  \begin{equation}
    \label{eq: defofoa}
    \oxn^{(1)} := \left\{ \omega: m(B_{[x-n,x+n],\omega}^{\bg,-})\leq e^{-\delta_1(2n+1)}\right\}\cap \widetilde{\Omega}_{x,n}.
  \end{equation}
  By Chebyshev's Inequality and Fubini's Theorem, we obtain part (1):
  \[
    \begin{split}
      \PP\left((\oxn^{(1)})^c\right)&\leq m \times \PP \left\{(\omega,z):(\omega,z)\in B_{[x-n,x+n]}^{\bg,-}, z\in\mathcal{I})\right\} + \PP(\widetilde{\Omega}_{x,n})\\
      &\leq m(\mathcal{I})e^{-(\eta_1-\delta)(2n+1)} + ne^{-\eta_1(2n+1)}.
    \end{split}
  \]
  Now for part (2), take any $\omega\in\oxn^{(1)}$, and any $(\gamma(z) - 2\epsilon,n,\omega,z)$-singular point $x$. By Lemma \ref{lemma: singular}, 
  \[
    (\omega,z)\in B_{[x-n,x+n],\epsilon}^{\beta,\gamma,-}\cup B_{[x-n,x-1],\epsilon}^{\beta,\cdot,+}\cup B_{[x+1,x+n],\epsilon}^{\cdot,\gamma,+},
  \]
  However, since $\omega\in\widetilde{\Omega}_{x,n}$, by Lemma \ref{lemma: uniformCS}, 
  \[
    (\omega,z) \notin B_{[x-n,x-1],\epsilon}^{\beta,\cdot,+}\cup B_{[x+1,x+n],\epsilon}^{\cdot,\gamma,+}.
  \]
  We see that $(\omega,z)\in B_{[x-n,x+n],\epsilon}^{\beta,\gamma,-}$. Thus 
  \[
    z\in B_{[x-n,x+n],\epsilon,\omega}^{\beta,\gamma,-}\quad \text{with}\quad  m(B_{[x-n,x+n],\epsilon,\omega}^{\beta,\gamma,-})\leq e^{-\delta(2n+1)},
  \]
  where the latter is due to \eqref{eq: defofoa}. Notice further that
  \[
    B_{[x-n,x+n],\epsilon,\omega}^{\beta,\gamma,-} = \{z: |P_{[x-n,x+n],\omega,z}^\bg|\leq e^{(\gamma(z)-\epsilon)(2n+1))}\}
  \]
  where for each $\omega$, $|P_{[x-n,x+n],\omega,z}^{\bg}|$ is a polynomial in $z$ with roots $\sigma(\ME_{[x-n,x+n],\omega}^\bg)$. Thus $B_{[x-n,x+n],\epsilon,\omega}^{\beta,\gamma,-}$ is a finite union of intervals, each centered around points of $\sigma(\ME_{[x-n,x+n],\omega}^\bg)$, of overall length less than $e^{-\delta(2n+1)}$. Thus,
  \[
    \text{dist}(z,\sigma(\ME_{[x-n,x+n],\omega}^\bg))\leq e^{-\delta(2n+1)}.
  \]
\end{proof}

\medskip

\subsection{The second set $\ob$}
\label{ss: Omega_3}
As mentioned above, the aim of choosing $\ob$ is to make sure $\sigma(\ME_{[x-n,x+n],\omega}^\bg)$ and $\sigma(\ME_{[x+n+1,x+3n+1],\omega}^\bg)$ are not exponentially close for $\omega\in\ob$.

\begin{lemma}
  \label{lemma: ob}
  For any $\delta>0$, there is $\eta_2(\delta)$, $N_2(\delta)$ s.t. for any $n>N_2$, $x\in\ZZ$, there is $\ob$, s.t. 
  \begin{enumerate}
    \item $\PP(\ob) \geq 1 - 2(2n+2)^3e^{-\eta_2(2n+1)}$,
    \item If $\omega \in \ob$, then 
    \[
      \operatorname{dist}\left(\sigma(\ME_{[x-n,x+n],\omega}^\bg),\sigma(\ME_{[x+n+1,x+3n+1],\omega}^\bg)\right) \geq 2e^{-\delta(2n+1)}
    \]
  \end{enumerate}
\end{lemma}

\begin{proof}
Since each entry in $\ME$ is bounded, there is $M$ s.t. \[
  |P_{[a,b],z}|\leq M^{b-a+1} , \quad \forall a\leq b\in\ZZ, \forall z\in\mathcal{I}.
\]
Choose $\epsilon'<\delta/2$. Apply Lemma \ref{lemma: LDT} to get  $\tilde{\eta}(\epsilon')$, $\tilde{N}_1(\epsilon')$. Choose $K\geq \frac{2\log M}{\delta - 2\epsilon'}$. Let $\eta_2 := \frac{\tilde{\eta}}{2K}$, $\tilde{N}_2 := K\tilde{N}_1$ and 
\[
  (\ob)^c := \bigcup\limits_{z_i\in Z(\omega)}\bigcup\limits_{(y_1,y_2)\in Y}\left(B_{[x-n,x+y_1-1],\epsilon',z_i}^{\beta,\cdot}\cup B_{[x+y_2+1,x+n],\epsilon',z_i}^{\cdot,\gamma}\right)\cup B_{[x-n,x+n],\epsilon',z_i}^{\bg}
\]
where 
\[
  \begin{split}
    &Y = \{(y_1,y_2): x-n\leq y_1\leq y_2\leq x+n, |y_1-(-n)|, |n-y_2|\geq \frac{n}{K}\},\\
    &Z = Z(\omega) = Z(\omega_{[x+n+1,x+3n+1]}) = \sigma(\ME_{[x+n+1,x+3n+1],\omega}).
  \end{split}
\]
We remark here that while $z_i(\omega)$ and $Z(\omega)$ depend on $\omega$,  they actually only depend on $\Omega_{[x+n+1,x+3n+1]}$ which is independent from $\Omega_{[x-n,x+n]}$. Thus $z_i = z_i(\omega) = z_i(\omega_{[x+n+1,x+3n+1]})$ in $B_{[x-n,x+n],z_i}^{\beta,\gamma}$ operates like any other fixed $z$ that does not depend on $\omega$. A rigorous argument is as follows:

For any fixed $\omega_c,\cdots,\omega_d $, with $[c,d]\cap[a,b]=\emptyset$, assume $d-c, b-a \geq \tilde{N}_1$. By independence,
\[
\mathbb{P}_{[c,d]^c}(B^{*}_{[a,b],\epsilon',z_{i,(\omega_c,\cdots,\omega_d)}})=\mathbb{P}_\ab (B^{*}_{[a,b],\epsilon',z_{i,(\omega_c,\cdots,\omega_d)}})\leq e^{-\eta_2(b-a+1)}
\]
where $*$ represents corresponding boundary conditions, $z_{i,(\omega_c,\cdots,\omega_d)}\in\sigma(\ME_{[c,d]})$. Applying to $[a,b]=[x-n,x+y_1-1]$ or $[x+y_2+1,x+n]$ or $[x-n,x+n]$, $[c,d]=[x+n+1,x+3n+1]$ and integrating over $\omega_{a},\cdots,\omega_{b}$, we obtain for $n\geq \tilde{N}_2$, 
\[
  \begin{split}
    &\mathbb{P}(B^{\beta,\cdot}_{[x-n,x+y_1-1],\epsilon',z_{i}}\cup B^{\cdot,\gamma}_{[x+y_2+1,x+n],\epsilon',z_{i}}) \leq 2e^{-\eta_2(\frac{n}{K}+1)},\\
    &\PP(B_{[x-n,x+n],\epsilon',z_i}^{\bg})\leq e^{-\eta_2(2n+1)}.
  \end{split}
\]
Thus we obtain part (1):
\[
  \PP(\ob) \geq 1 - (2n+1)((2n+1)^2+1)2e^{-\eta_2\frac{n}{K}} \geq 1 - 2(2n+2)^3e^{-\eta_2\frac{n}{K}}
\]

We prove part (2) by contradiction. Let $\omega\in\ob$, assume that there is $z_i\in\sigma(\ME_{[x+n+1,x+3n+1]})$, $z_j\in \sigma(\ME_{[x-n,x+n]})$ s.t. 
\[
  |z_i - z_j|\leq 2e^{-\delta(2n+1)}.
\] 
Then 
\[
  \Vert G_{[x-n,x+n],\omega, z_i}^\bg\Vert \geq \frac{1}{2}e^{\delta(2n+1)}.
\]
Thus there are $x-n\leq y_1\leq y_2\leq x+n$ s.t. 
\[
  \frac{|P_{[x-n,x+y_1-1],\omega,z_i}P_{[x+y_2+1,x+n],\omega,z_i}|}{|P_{[x-n,x+n],\omega,z_i}|} = |G_{[x-n,x+n],\omega, z_i}^\bg(y_1,y_2)|\geq \frac{1}{2n}e^{\delta(2n+1)}.
\]
There are three cases, and we claim that each leads to a contradiction.
\begin{enumerate}
  \item If $|y_1 - (-n)|\geq \frac{n}{K}$, $|n-y_2|\geq \frac{n}{K}$, since 
  \[
    \omega\notin B_{[x-n,x+y_1-1],\epsilon',z_i}^{\beta,\cdot}\cup B_{[x+y_2+1,x+n],\epsilon',z_i}^{\cdot,\gamma}\cup B_{[x-n,x+n],\epsilon',z_i}^\bg,
  \]
   if $K>1$, we have 
  \[
    \frac{1}{2n}e^{\delta(2n+1)} \leq e^{(\gamma(z_i) + \epsilon')\frac{2n}{K} - (\gamma(z_i) - \epsilon')(2n+1)}\leq e^{(2n+1)(2\epsilon')}.
  \]
   But $\delta>2\epsilon'$. Thus when $n$ is large enough, say, $n>\tilde{N}_3$, there will be a contradiction.
   \item If one of $|y_1 - (-n)|$ and $|n-y_2|\geq \frac{n}{K}$, then if $K>1$, we have 
   \[
     \frac{1}{2n}e^{\delta(2n+1)} \leq M^{\frac{n}{K}}e^{n\frac{(\gamma(z_i) + \epsilon')}{K} - (\gamma(z_i) - \epsilon')(2n+1)}\leq e^{(2n+1)(\frac{\log M}{2K}+2\epsilon')}
   \]
   By our choice of $K\geq \frac{2\log M}{\delta - 2\epsilon'}$, we have $\delta > \frac{\log M}{2K} + 2\epsilon'$. Thus again, when $n$ is large enough, say, $n>\tilde{N}_4$, we arrive at a contradiction.
   \item If both $|y_1 - (-n)|\leq \frac{n}{K}$, $|n-y_2|\leq \frac{n}{K}$, then 
   \[
     \frac{1}{2n}e^{\delta(2n+1)}\leq M^{\frac{2n}{K}}e^{-(\gamma(z_i) - \epsilon')(2n+1)}\leq e^{(2n+1)(\frac{\log M}{2K} + \epsilon')}
   \]
   By our choice of $K$, we have $\delta>\frac{\log M}{2K} +\epsilon'$. Thus when $n$ is large enough, say, $n>\tilde{N}_5$, again we arrive at a contradiction.
\end{enumerate}
Take $N_2 = \max\{\tilde{N}_1,\tilde{N}_2,\tilde{N}_3,\tilde{N}_4,\tilde{N}_5\}$. Then for any $n>N_2$, we have a contradiction for all three cases, and hence
\[
  \text{dist}\left(\sigma(\ME_{[x-n,x+n],\omega}^\bg),\sigma(\ME_{[x+n+1,x+3n+1],\omega}^\bg)\right) \geq 2e^{-\delta(2n+1)}
\]
\end{proof}
We now prove Theorem \ref{thm: regular}:
\begin{proof}[Proof of Theorem \ref{thm: regular}]
By Lemma \ref{lemma: oa}, for any $\epsilon>0$, we can find $\eta_1(\epsilon), N_1(\epsilon)$ and $\delta = \eta_1/2$, s.t. (1) and (2) of Lemma \ref{lemma: oa} hold.  For such $\delta$, apply Lemma \ref{lemma: ob} to find $\eta_2$, $N_2$ and $\ob$ for any $x\in\ZZ$, $n>N_2$. Now let $\eta := \min\{\eta_1,\eta_2/2\}$, $N := \max\{N_1,N_2\}$. Set
\[
  \oxn := \oa\cap \Omega_{x+2n+1,n}^{(1)} \cap \ob.
\]
Then we obtain part (1):
\[
  \begin{split}
    \PP(\oxn) &\geq 1 - 2m(\mathcal{I})e^{-\eta_1(2n+1)/2} - 2ne^{-\eta_1(2n+1)} - 2(2n+2)^3e^{-\eta_2(2n+1)}\\
    &\geq 1 - Ce^{-\eta(2n+1)}.
  \end{split}
\]
As for part (2), let $\omega\in\oxn$. Assume both $x$ and $x+2n+1$ are $(\gamma(z_i) - 2\epsilon,n,\omega,z)$-singular. Then by Lemma \ref{lemma: oa}, we have 
\[
  \begin{split}
    &\text{dist}(z,\sigma(\ME_{[x-n,x+n],\omega}))\leq e^{-\delta(2n+1)},\\
    &\text{dist}(z,\sigma(\ME_{[x+n+1,x+3x+1],\omega}))\leq e^{-\delta(2n+1)}.
  \end{split}
\]
Thus \[
  \text{dist}\left(\sigma(\ME_{[x-n,x+n],\omega}^\bg),\sigma(\ME_{[x+n+1,x+3n+1],\omega}^\bg)\right) \leq 2e^{-\delta(2n+1)}.
\]
However, Lemma \ref{lemma: ob} guarantees that if $\omega\in \oxn$, then 
\[
  \text{dist}\left(\sigma(\ME_{[x-n,x+n],\omega}^\bg),\sigma(\ME_{[x+n+1,x+3n+1],\omega}^\bg)\right) > 2e^{-\delta(2n+1)}.
\]
which is a contradiction. Thus at least one of the two points $x$ or $x+2n+1$ must be regular. 
\end{proof}

\appendix

\section{}
\label{App.A}
\begin{lemma}\label{matrixA}
 The matrix $S_z=(z(\ML)^*-\MM)$ is tridiagonal. We omit $z$ and denote the $(i,j)$-entry by $A_{i,j}$ for convenience, when $a\leq i,j\leq b$. Then we have
  \[
   A_{j,j} =
   \begin{cases}
     z\alpha_j+\alpha_{j-1}, \quad &j~even,\\
     -z\overline{\alpha_{j-1}}-\overline{\alpha_j},\quad & j~odd,\\
   \end{cases},\quad
   A_{j+1,j}=A_{j,j+1}=
   \begin{cases}
     z\rho_j,\quad &j~even,\\
     -\rho_j,\quad &j~odd.\\
   \end{cases}
  \]
\end{lemma}
\begin{remark}\label{remark Abg}
  If we modify the extended CMV matrix at $a-1$ and $b$ by $\beta$ and $\gamma$, then the corresponding matrix $A_\abz ^{\beta,\gamma}$ is the restriction of $S_z$ on $[a,b]$ but with $\alpha_{a-1}=\beta$ and $\alpha_b=\gamma$. Fix an interval $[a,b]$. We denote $S_z^{\beta,\gamma}$ by the infinite matrix $S_z$ with  $\alpha_{a-1}=\beta$ and $\alpha_b=\gamma$.
\end{remark}
\begin{lemma}\label{lemma: poisson}
  Let $\Psi$ solve $\ME\Psi=z\Psi$, then for $a<n<b$,
\[
  \begin{split}
    \Psi(n)=&-G_{[a,b],z}^{\beta,\gamma}(n,a)
    \begin{cases}
      \Psi(a)(z\bar{\beta}-z\bar\alpha_{a-1})+\Psi(a-1)z\rho_{a-1},\qquad &a~odd\\
      \Psi(a)(\alpha_{a-1}-\beta)-\Psi(a-1)\rho_{a-1},\qquad &a~even\\
    \end{cases}\\
    &-G_{[a,b],z}^{\beta,\gamma}(n,b)
    \begin{cases}
      \Psi(b)(-\bar\alpha_b+\bar\gamma)-\Psi(b+1)\rho_b,\qquad &b~odd,\\
      \Psi(b)(z\alpha_b-z\gamma)+\Psi(b+1)z\rho_b,\qquad &b~even
    \end{cases}
    \end{split}
\]
\end{lemma}
\begin{proof}
  \[
  \begin{split}
    &P_\ab S_z\Psi=0\\
    \Rightarrow &P_\ab S_z^{\beta,\gamma}\Psi+P_\ab (A-A^{\beta,\gamma})\Psi=0\\
    \Rightarrow &P_\ab S_z^{\beta,\gamma}(P_\ab \Psi+P_{[a,b]^c}\Psi)+P_\ab (S_z-S_z^{\beta,\gamma})\Psi=0\\
    \Rightarrow &A_\abz ^{\beta,\gamma}\Psi=-P_\ab (S_z-S_z^{\beta,\gamma})\Psi-P_\ab S_z^{\beta,\gamma}P_{[a,b]^c}\Psi
\end{split}
\]
There are two terms on the right hand side. One is always $0$:
\[
\begin{split}
  &P_\ab S_z^\bg P_{[a,b]^c}\Psi=\\
  &\begin{bmatrix}
   \ddots &\ddots \\
 A^{\beta,\gamma}_{a,a-1} & A^{\beta,\gamma}_{a,a}  & A^{\beta,\gamma}_{a,a+1} & 0 & \cdots \\
  \\
  & \ddots & \ddots & \ddots & \\
  \\
\cdots & 0 & A^{\beta,\gamma}_{b,b-1} & A^{\beta,\gamma}_{b,b} & A^{\beta,\gamma}_{b,b+1} & \\
& & & \ddots & \ddots\\
\end{bmatrix}
\begin{bmatrix}
  \Psi(a-1)\vspace{1.7mm}\\
  0\vspace{1.2mm}\\
  \\
\vdots  \\
  \\
  0\vspace{1.7mm}\\
  \Psi(b+1)\vspace{1.2mm}\\
\end{bmatrix}
=\begin{bmatrix}
0\vspace{1.7mm}\\
0\vspace{1.2mm}\\
\\
\vdots\\
\\
0\vspace{1.7mm}\\
0\vspace{1.2mm}\\
\end{bmatrix}
\begin{matrix}
  row~a-1\vspace{1.7mm}\\
  row~a\vspace{1.2mm}\\
  \\
  \vdots\\
  \\
  row~b\vspace{1.7mm}\\
  row~b+1\vspace{1.2mm}\\
\end{matrix},
\\
\end{split}
\]
 where we use Remark \ref{remark Abg} and Lemma \ref{matrixA} to get $A_{a,a-1}^{\beta,\gamma}=A_{b,b+1}^{\beta,\gamma}=0$. 
 
 The other has two non-zero terms:
 \[
 \begin{split}
     &P_\ab (S_z-S_z^{\beta,\gamma})\Psi=\\
     &\begin{bmatrix}
     \ddots &\ddots \\
     A^-_{a,a-1} & A^-_{a,a}  & A^-_{a,a+1} & 0 & \cdots \\
     \\
     & \ddots & \ddots & \ddots &\\
      \\
      \cdots & 0 & A^-_{b,b-1} & A^-_{b,b} & A^-_{b,b+1} & \\
      & & & \ddots & \ddots\\
\end{bmatrix}
\begin{bmatrix}
  \Psi(a-1)\vspace{1.7mm}\\
  \Psi(a)\vspace{1.2mm}\\
  \\
\vdots  \\
  \\
  \Psi(b)\vspace{1.7mm}\\
  \Psi(b+1)\vspace{1.2mm}\\
\end{bmatrix}
=\begin{bmatrix}
\vdots\vspace{1.7mm}\\
\Psi(a-1)A_{a,a-1}^-+\Psi(a)A_{a,a}^-\vspace{1.2mm}\\
\\
\cdots\\
\\
\Psi(b)A_{b,b}^-+\Psi(b+1)A_{b,b+1}^-\vspace{1.7mm}\\
\vdots\vspace{1.2mm}\\
\end{bmatrix}
 \end{split}
 \]
 where $A_{x,y}^-=A_{x,y}-A^{\beta,\gamma}_{x,y}$. 
 
 Now 
 \[
   \Psi(n) = -G_\abz^\bg(n,b)\left( \Psi(b)A_{b,b}^-+\Psi(b+1)A_{b,b+1}^- \right) - G_\abz^\bg \left(\Psi(a-1)A_{a,a-1}^-+\Psi(a)A_{a,a}^- \right)
 \]
 where $A^-_{x,y}$ is derived from Lemma \ref{matrixA} and Remark \ref{remark Abg}
 \[
 A_{a,a-1}^-=
 \begin{cases}
   z\rho_{a-1},\qquad a~odd,\\
   -\rho_{a-1},\qquad a~even.\\
 \end{cases}\qquad
 A^-_{a,a}=
 \begin{cases}
   -z\bar\alpha_{a-1}+z\bar\beta,\qquad a~odd,\\
   \alpha_{a-1}-\beta,\qquad a~even.
 \end{cases}
 \]
 \[
 A^-_{b,b+1}=
 \begin{cases}
   -\rho_b, b~odd,\\
   z\rho_b,b~even.\\
 \end{cases}\qquad
 A^-_{b,b}=
 \begin{cases}
   -\bar\alpha_b+\bar\gamma,\qquad b~odd,\\
   z\alpha_b-z\gamma,\qquad b~even.\\
 \end{cases}.
\]
That proves the result.
\end{proof}

\section{Corrections}
\label{App.B}
As mentioned in the introduction, we provide  corrections for some of the mistakes from \cite{Kruger13}, \cite{OPUC05}, \cite{7Authors19}, \cite{Onefoot05}. We first provide the correct results in their notations and then, for the reader's convenience, we rewrite them in our notation when there is a correspondence. Finally, we give either a short proof or a reference for those citations in \cite{Kruger13} which are invalid now.

\subsection{Corrections for \cite{Kruger13}}
  \begin{enumerate}
    \item Formula (3.6) in Lemma 3.3 should be $\mathcal{C} = \mathcal{E}_{-1,\cdot}^{[0,\infty)}$. Or in our notation, $\mathcal{C}=\ME_{[0,+\infty]}^{-1,\cdot}$. It follows from the definition, see Remark \ref{rmk: oneside}.
    \item Formula (3.14) in Lemma 3.6 should be $\Phi_n(z) = \Phi_{-1,\cdot}^{[0,n-1]}(z)$. Or in our notation, $\Phi_n(z) = \MP_{[0,n-1]}^{-1,\cdot}(z)$. See \cite[Theorem 5.3]{Onefoot05} for a proof.
    \item Formula (3.16) and (3.17) in Lemma 3.7 should be 
    \begin{equation}
      \label{eq: boundary_to_inside}
      \Phi_n^\beta(z) = \Phi_{-\bar{\beta},\cdot}^{[0,n-1]}(z)\quad \text{and}\quad\Phi_n^\beta(z;\gamma) = \Phi_{-\bar{\beta},\gamma}^{[0,n-1]}(z)
    \end{equation}
     where $\Phi_n^\beta(z;\gamma)$ means first replacing $\alpha_{n-1}$ by $\gamma$, then multiplying every $\alpha_{k}, 0\leq k <n-1$ and $\gamma$ by $\beta$ (instead of the reversed order). In our notation there is no direct correspondence, but if we denote $X_{[a,b]}^\bg(\zeta)$ to be $X_{[a,b]}^\bg$ with all coefficients $\alpha_a,\cdots,\alpha_{b-1},\gamma$ being multiplied by $\zeta$, where $X$ can be $\MC,\ME,\MP,\cdots$, then
    \[
      \MP_{[0,n-1]}^{-1,\cdot}(\beta) = \MP_{[0,n-1]}^{-\bar{\beta},\cdot} \quad \text{and}\quad \MP_{[0,n-1]}^{-1,\gamma}(\beta) = \MP_{[0,n-1]}^{-\bar{\beta},\gamma}.
    \]
    See \cite[Theorem 4.2.9]{OPUC05} for a proof. See also \cite[Theorem 5.6]{Onefoot05} for a clear restatement but with a typo: If $D$ is a diagonal matrix with elements $1,\lambda^{-1},1,\lambda^{-1},\cdots$, and $\mathcal{M}_\lambda$ differs from $\mathcal{M}$ by having $\lambda$ in the $(0,0)$-position instead of $1$, then $D\MC({\lambda\alpha})D^{-1} = \mathcal{L}(\{\alpha_n\})\mathcal{M}_{\bar{\lambda}}(\{\alpha_n\})$.
    \item Formula (3.18) in Prop. 3.8 should be 
    \[
      |G_\bg^{[a,b]}(z;k,l)| = \frac{1}{\rho_l}\left\vert \frac{\phi_{\beta,\cdot}^{[a,k-1]}(z)\phi_{\cdot, \gamma}^{[l+1,b]}(z)}{\phi_\bg^{[a,b]}(z)}\right\vert.
    \]
    In our notation, the equality is given in \eqref{eq: GreenFunc}. Notice that we have no extra parameters $\frac{1}{\rho_l}$ because our definition of $P_\aboz^\bg$ is different from the corresponding definition of $\phi_\bg^{[a,b]}(z)$. This result follows by direct computation using Cramer's rule.
    \item Formula (3.22) in Lemma 3.10 should be 
    \[
      T_{[a,b]}(z) = \frac{1}{2}\begin{pmatrix}
        \phi_{-1,\cdot}^{[a,b]}(z)+ \phi_{1,\cdot}^{[a,b]}(z) & \phi_{-1,\cdot}^{[a,b]}(z) - \phi_{1,\cdot}^{[a,b]}(z)\\
        (\phi_{-1,\cdot}^{[a,b]})^*(z)  - (\phi_{1,\cdot}^{[a,b]})^*(z) &(\phi_{-1,\cdot}^{[a,b]})^*(z) +(\phi_{1,\cdot}^{[a,b]})^*(z)
      \end{pmatrix}.
    \]
    where we used a different formula for $T_{[a,b]}$, i.e. \eqref{*}. For a proof of the correct form, see \cite[(3.2.17), (3.2.27)]{OPUC05}.
    \item Formula (3.23), (3.24) in Cor. 3.11 should be
    \[
      \begin{pmatrix}
        \phi_{\beta, \cdot}^{[a,b]}(z)\\
        -\beta(\phi_{\beta, \cdot}^{[a,b]})^*(z)
      \end{pmatrix} = T^{[a,b]}(z) \begin{pmatrix}
        1\\ -\beta
      \end{pmatrix}
    \]
    and
    \[
      \phi_\bg^{[a,b]}(z) = \frac{1}{\rho_b}\left \langle \begin{pmatrix}
        z\\ \beta\bar{\gamma}
      \end{pmatrix}, T_{[a,b-1]} \begin{pmatrix}
        1 \\ -\beta
      \end{pmatrix}\right\rangle.
    \]
    Note in the proof of Cor. 3.11, they used (3.2.26) in \cite{OPUC05}, which has a typo and the correct form should be 
    \begin{equation}
      \label{eq: initial_boundary}
      \begin{pmatrix}
        \phi^\lambda_{n+1}\\ \bar{\lambda}(\phi_{n+1}^\lambda)^*
      \end{pmatrix} = T_n(z)\begin{pmatrix}
        1\\\bar{\lambda}
      \end{pmatrix}.
    \end{equation}
    \begin{proof}
      The first equality follows from \eqref{eq: boundary_to_inside} and \eqref{eq: initial_boundary}. The second equality follows from the first equality and 
      \[
        \psi^{[a,b]}_{\beta,\gamma} = \Phi_n^{-\bar{\beta}}(z;\gamma) = \frac{1}{\rho_b}(\Phi^{[a,b-1]}_{\beta,\cdot}z+\beta\bar{\gamma}(\phi^{[a,b-1]}_{\beta,\cdot})^*).
      \]
    \end{proof}
  \end{enumerate}

  \subsection{Corrections for \cite{7Authors19}}
  \begin{enumerate}
    \item Equation (7.4) should be 
    \[
      \phi_{\omega,[a,b]}^{\beta,\gamma}(z) = \frac{1}{\rho_b}\left \langle \begin{pmatrix}
        z\\ \beta\bar{\gamma}
      \end{pmatrix}, S_{b-a}^z(T^a\omega) \begin{pmatrix}
        1 \\ -\beta
      \end{pmatrix}\right\rangle.
    \]
    We believe this can  be used then to derive \eqref{eq: LDT3} by \cite[Theorem 5]{Tsay99}. Then one could complete the proof of double elimination in \cite{7Authors19}.
    \item The second to the last equation of Page 39 should be 
    \[
      |G_{\omega,\Lambda}^{\tau_1,\tau_2}(j,k;z)| = \left\vert \frac{\phi_{\omega,[a,j-1]}^{\tau_1,\cdot}(z)\phi_{\omega,[k+1,b]}^{\cdot,\tau_2}(z)}{\rho_{\omega,[a,b]}^{\tau_1,\tau_2}(z)}\right\vert \prod\limits_{i = j}^{k-1}\rho_i.
    \]
  \end{enumerate}

\section*{Acknowledgement}
This research is partially supported by NSF DMS-1901462 and DMS-2052899. I am very grateful to my advisor Svetlana Jitomirskaya for her generous support and patience and helpful
suggestions on this paper. I also want to thank Simon Becker for his moral support during the completion of the paper.

\bibliographystyle{unsrt}
\bibliography{mybib}

\end{document}

%% file: settings/settings.tex
\usepackage{amsfonts}
 \usepackage{mathabx}
 \usepackage{cases}
 \usepackage[numbers,sort&compress]{natbib}
 \usepackage{hyperref}
 \usepackage[usenames,dvipsnames]{color}

\newtheorem{thm}{Theorem}
\newtheorem{lemma}{Lemma}[section]

\newtheorem{prop}{Proposition}
\theoremstyle{definition}
\newtheorem{definition}{Definition}

\theoremstyle{remark}
\newtheorem{remark}{Remark}

\numberwithin{equation}{section}

\newcommand{\DD}{\mathbb{D}}
\newcommand{\ZZ}{\mathbb{Z}}
\newcommand{\EE}{\mathbb{E}}
\newcommand{\RR}{\mathbb{R}}
\newcommand{\CC}{\mathbb{C}}
\newcommand{\PP}{\mathbb{P}}
\newcommand{\NN}{\mathbb{N}}
\newcommand{\ME}{\mathcal{E}}
\newcommand{\MC}{\mathcal{C}}
\newcommand{\MP}{\mathcal{P}}
\newcommand{\ML}{\mathcal{L}}
\newcommand{\MM}{\mathcal{M}}
\newcommand{\ab}{{[a,b]}}
\newcommand{\abz}{{[a,b],z}}
\newcommand{\abo}{{[a,b],\omega}}
\newcommand{\aboz}{{[a,b],\omega,z}}
\newcommand{\bg}{{\beta,\gamma}}
\newcommand{\oxn}{\Omega_{x,n}}
\newcommand{\oa}{\Omega_{x,n}^{(1)}}
\newcommand{\ob}{\Omega_{x,n}^{(2)}}

